\pdfoutput=1
\documentclass[aps,prd,showpacs,preprintnumbers,reprint]{revtex4-1}

\usepackage[colorlinks=true, pdfstartview=FitV, linkcolor=blue, citecolor=blue, urlcolor=blue]{hyperref}
\usepackage{graphicx}
\usepackage{amsmath}
\usepackage{amssymb}
\usepackage{amsfonts}
\usepackage{bm}
\usepackage{bbold}
\usepackage{color}

\DeclareMathOperator\ee{e}
\DeclareMathOperator\Tr{Tr}

\newcommand{\ckakko}[1]{\left\{#1\right\}}
\newcommand{\mkakko}[1]{\left({#1}\right)}
\newcommand{\1}{\mathbb{1}}
\newcommand{\der}{\partial}
\newcommand{\lat}{\mathrm{lat}}
\newcommand{\ad}{\mathrm{ad}}
\newcommand{\YM}{\mathrm{YM}}
\newcommand{\dd}{\mathrm{d}}

\newcommand{\SU}{\text{SU}}
\newcommand{\U}{\text{U}}

\newcommand{\calO}{\mathcal{O}}
\renewcommand{\epsilon}{\varepsilon}
\renewcommand{\bar}[1]{\overline{#1}} 
\newcommand{\muu}{\hat{\mu}}

\newcommand{\PP}{\mathfrak{P}}

\begin{document}
\preprint{RIKEN-QHP-169}
\title{Asymptotically free lattice gauge theory in five dimensions}

\author{Takuya Kanazawa}
\affiliation{iTHES Research Group and Quantum Hadron Physics 
Laboratory, RIKEN, Saitama 351-0198, Japan}

\author{Arata~Yamamoto}
\affiliation{
Department of Physics, The University of Tokyo, Tokyo 113-0033, Japan
}
\affiliation{
Theoretical Research Division, Nishina Center, RIKEN, Saitama 351-0198, Japan
}

\date{\today}
\allowdisplaybreaks
\begin{abstract}
A lattice formulation of
Lifshitz-type gauge theories is presented. 
While the Lorentz-invariant Yang-Mills theory is not renormalizable 
in five dimensions, non-Abelian
Lifshitz-type gauge theories are renormalizable and asymptotically free. 
We construct a lattice gauge action and numerically examine the continuum limit and the bulk phase structure.
\end{abstract}

\maketitle

\section{Introduction}

Since olden times, the Lifshitz-type anisotropic field theory \cite{Lifshitz1,Lifshitz2} has been considered in various condensed matter systems.
In recent years, the Ho\v{r}ava-Lifshitz-type gravity \cite{Horava:2009uw} has received much interest. 
Its analogues in non-gravitational quantum field theories have also been 
discussed intensively \cite{Anselmi:2008bq,Anselmi:2008bs,Horava:2008jf,Visser:2009fg,
Dijkgraaf:2009gr,Anselmi:2009vz,Chen:2009ka,Dhar:2009dx,Das:2009ba,Iengo:2009ix,
Kawamura:2009re,Orlando:2009az,Kaneta:2009ci,Das:2009fb,
Alexandre:2009sy,Andreev:2009ba,Chao:2009dw,Dhar:2009am,
Iengo:2010xg,Romero:2010tc,Anagnostopoulos:2010gw,
Kobakhidze:2010cq,Bakas:2011nq,Hatanaka:2011kg,Eune:2011zw,
He:2011hb,Gomes:2011pq,Farakos:2011px,Bakas:2011uf,
Kikuchi:2011np,Farias:2011aa,Gomes:2011di,Farakos:2011cg,
Farakos:2012cs,Montani:2012ve,Farias:2012rd,
Lozano:2012aw,Adam:2012dg,Alves:2013xda,Gomes:2013jba,
Alexandre:2013wua,Hoyos:2013qna,Farias:2013rya,Arav:2014goa}. 
Besides a purely theoretical interest on its own, there are several motivations to look into such non-Lorentz invariant 
field theories in the context of physics beyond the Standard Model. Firstly, various extradimensional models have been 
proposed in attempts to remedy the hierarchy problem in particle physics, and 
their common problem is that gauge theories in higher dimensions are usually unrenormalizable and need a UV cutoff scale. 
In anisotropic Lifshitz-type theories with higher derivative terms, the behavior of propagators in UV is improved and 
one can construct renormalizable theories in higher dimensions, which may be appreciated as UV completion of 
phenomenologically introduced extradimensional models. In addition, such renormalizable theories admit four-fermion 
interactions, which may shed new light on the traditional technicolor models in which Higgs particle is generated 
from strong-coupling dynamics of fermions. We refer the reader to \cite{Alexandre:2011kr} for a review on these directions.

We note that anisotropic gauge theories are also expected to arise as an effective theory at quantum critical points in certain condensed matter systems, see \cite{Hermele2004,Moessner2003, Ardonne:2003wa, Freedman:2004ki,Watanabe:2014qla} and references therein. 
Cold atomic gases may also provide a venue for non-Abelian gauge theories \cite{Banerjee:2012xg,2013NatCo...4E2615T,Zohar:2012xf}. 

In this work we propose a lattice formulation of an anisotropic non-Abelian gauge theory put forward 
by Ho\v{r}ava \cite{Horava:2008jf}.
The action of this Ho\v{r}ava-Lifshitz-type gauge 
theory in $(1+D)$-dimensional Euclidean spacetime reads 
\begin{equation}
  \begin{split}
    S &= \frac{1}{2} \int \dd x_0 \dd^Dx \bigg[ \frac{1}{e^2}\Tr(E_i E_i) 
    \\
    &\quad + \frac{1}{g^2}\Tr\ckakko{ (D_i^{\ad}F_{ik})(D_j^{\ad}F_{jk}) } \bigg]\,,
    \label{eq:action}
  \end{split}
\end{equation}
where the indices $i,j,k$ run from $1$ to $D$, and 
\begin{subequations}
\begin{alignat}{2}
  & E_i = F_{0i} \\
  & F_{ij} = -i[D_i,D_j] = \der_i A_j - \der_j A_i + i[A_i, A_j] \\
  & D_i = \der_i + iA_i \\
  & D_i^{\ad}F = \der_i F + i[A_i, F]\,. 
\end{alignat}
\end{subequations}
The gauge field $A_i\equiv A_i^a T^a$ takes values in the Lie algebra 
of a non-Abelian compact Lie group. 
For the second term of \eqref{eq:action} to be nonzero, $d\equiv1+D\geq 3$ is required. 
There are two couplings, $e^2$ and $g^2$. 
In a weighted power counting with the dimensions of fields 
$[A_0]=2$ and $[A_i]=1$, we find $[e^2]=[g^2]=4-D$. The critical dimension is 
$d=1+4$, for which the couplings are marginal. 
According to a general rule \cite{Anselmi:2008bq,Anselmi:2008bs}, 
renormalizability demands that all terms with weighted dimensions less than or equal to $D+2$ 
(such as $\Tr(F_{ij}F_{jk}F_{ki})$ and 
$\Tr\big\{ (D_i^{\ad}F_{jk})(D_i^{\ad}F_{jk}) \big\}$) be retained in the action. 
Nevertheless it was argued by Ho\v{r}ava that for $d=5$ the theory \eqref{eq:action} is 
renormalizable and asymptotically free \cite{Horava:2008jf}. This remarkable property is a consequence of  
the fact that the action \eqref{eq:action} satisfies the so-called \emph{detailed balance condition}; 
that is to say, the spatial part of the anisotropic action in $d$ dimensions consists of a square of the equation of motion of 
a theory living in $d-1$ dimensions. This particular form of anisotropic actions is known to arise 
in the Fokker-Planck dynamics of stochastic quantization \cite{Parisi:1980ys}, where a fictitious fifth dimension is introduced 
as a device of quantization. When this condition is met, the renormalization property of a theory is 
greatly simplified thanks to a special BRS-type symmetry \cite{ZinnJustin:1986eq}. Borrowing results from  
perturbative calculations for stochastic quantization of Yang-Mills theory \cite{Bern:1986jj,Okano:1986vr}, 
Ho\v{r}ava showed that the theory \eqref{eq:action} for $d=5$ is renormalizable and asymptotically free. 

While renormalizability in the continuum requires $d\leq 5$, we will shortly see that 
the theory can be discretized on a lattice in any $d\geq 3$ dimensions, thus opening a way toward a 
non-perturbative study of Ho\v{r}ava-Lifshitz-type gauge theories.  
With a soft deformation term, the theory restores effective Lorentz invariance in the infrared \cite{Horava:2008jf}, hence 
the theory may be considered as a UV completion of the non-renormalizable Yang-Mills theory 
in five dimensions \cite{Creutz:1979dw,Dimopoulos:2006qz,Irges:2006hg,Irges:2009qp,deForcrand:2010be,Farakos:2010ie,Knechtli:2011gq,DelDebbio:2012mr,
DelDebbio:2013rka,Itou:2014iya}. 

This paper is structured as follows. In Section \ref{sc:latticeformulation} 
we present a lattice action for the Ho\v{r}ava-Lifshitz-type 
gauge theory and discuss its continuum limit. 
In Section \ref{sc:numerics} the setup of our 
lattice simulation is outlined and the first numerical results of this theory 
for the $\SU(3)$ gauge group are presented. 
Section \ref{sc:summary} is devoted to summary and conclusions. 
Some technical details on the classical continuum limit are presented in appendix \ref{sc:continuum}.
Lattice actions for more general terms in the continuum are discussed in appendix \ref{sc:app2}.

\section{Lattice formulation}
\label{sc:latticeformulation}

In the following, for convenience, we call the isotropic $D$ dimensions 
``space'' and the other one dimension ``time'' although it is not necessarily so.
The spatial lattice spacing is denoted by $a$ and the temporal lattice spacing by $b$.
The mass dimensions are $[a]=-1$ and $[b]=-2$ according to the standard 
weighted power counting for Lifshitz-type theories \cite{Alexandre:2011kr}. 
Unit vectors in $x^\mu$ direction 
will be denoted as $\muu$ for $\mu=0,1,\dots,D$. 

The temporal and spatial link variables are defined as
$U_0(x) \equiv \mathrm{P}\exp\mkakko{ i\int_{x}^{x+b\hat{0}}\dd y\,A_0(y) } 
\simeq \exp(ibA_0(x))$ and $U_i(x) \equiv 
\mathrm{P}\exp\mkakko{ i\int_{x}^{x+a\hat{i}}\dd y\,A_i(y) } \simeq 
\exp(iaA_i(x))$, respectively.

\begin{figure}[t]
  \centering \includegraphics[width=5cm]{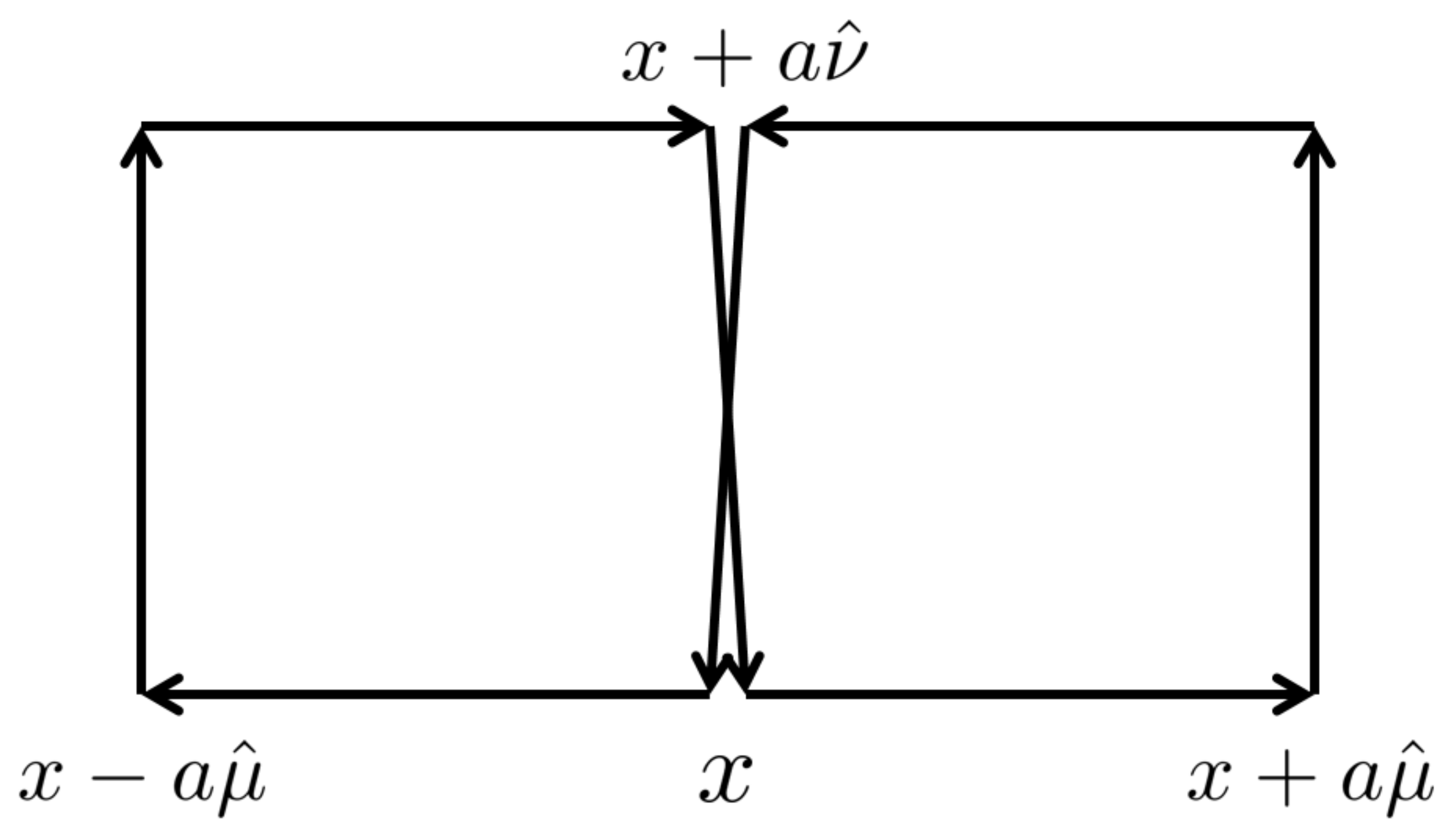}
  \caption{\label{fg:Wloop}
  A $2\times 1$ twisted Wilson loop $T_{\mu\nu}(x)$.
  }
\end{figure}

We define the lattice Ho\v{r}ava-Lifshitz gauge theory as
\begin{equation}
  Z=\int \mathcal{D}U\,\exp(-S_\lat)
\end{equation}
with
\begin{equation}
  \begin{split}
    S_\lat
    & \equiv \frac{1}{e_\lat^2} \sum_{x}\sum_{i=1}^{D}\mathrm{Re} \Tr \Big\{\1 - P_{0i}(x) \Big\} 
    \\
    &\quad + \frac{1}{g_\lat^2} \sum_{x}\sum_{j=1}^{D}
    \mathrm{Re}\Tr \bigg\{ \1 - \prod_{\substack{i=1\\ i\ne j}}^{D}T_{ij}(x) \bigg\} \,, 
    \label{eq:Slat}
  \end{split}
\end{equation} 
where $\1$ denotes the unit matrix. 
The temporal component of $S_{\lat}$ 
includes a $1\times 1$ plaquette $P_{\mu\nu}(x)$, which is well known 
in the lattice Yang-Mills theory, while the spatial component of $S_{\lat}$ 
includes a $2\times 1$ twisted loop $T_{\mu\nu}(x)$, which is shown in Fig.~\ref{fg:Wloop}. 
Such a rectangular loop has been considered for improved lattice actions \cite{Luscher:1984xn}. 
We remark that the ordering of $T$'s in the product $\prod T_{\mu\nu}(x)$ is inessential, 
because as we shall shortly see, only subleading terms irrelevant 
in the continuum limit are affected by this ordering. Note also that 
gauge invariance is maintained, since 
all the twisted loops begin and end at the same point $x$.

We can check the naive continuum limit of this lattice action using 
the Baker-Campbell-Hausdorff (BCH) formula.
The temporal plaquette may be evaluated as
\begin{equation}
  P_{0i}(x) = \exp\Big( iab F_{0i}(x) + \calO(a^2b,ab^2) \Big) \,. 
\end{equation}
Hence
\begin{equation}
\begin{split}
&  \sum_{x}\sum_{i=1}^{D}\mathrm{Re}\Tr\big\{ \1 - P_{0i}(x) \big\} \\
&= \frac{a^2b^2}{2}\sum_{x}\sum_{i=1}^{D}\Tr\ckakko{ F_{0i}(x)^2 } + \calO(a^3b^2,a^2b^3) .
\label{eq:Pact2}
\end{split}
\end{equation}
Next, the twisted loop is given (cf.~appendix \ref{sc:continuum}) by 
\begin{align}
   T_{ij}(x) = \exp \big( ia^3 D^{\ad}_i F_{ij}(x) + \calO(a^4) \big)\,. 
   \label{eq:Tij}
\end{align}
Then
\begin{align}
    & \quad \sum_{x}\sum_{j=1}^{D}\mathrm{Re}\Tr\bigg\{ \1- \prod_{\substack{i=1\\ i\ne j}}^{D}T_{ij}(x) \bigg\} 
    \notag
    \\
    & = \sum_{x}\sum_{j=1}^{D}\mathrm{Re}\Tr\bigg\{ \1 - \exp\bigg( ia^3 \sum_{i=1}^{D} D^{\ad}_i F_{ij}(x) + \calO(a^4) \bigg) \bigg\} 
    \notag
    \\
    & = \frac{a^6}{2}\sum_{x}\sum_{j=1}^{D} \Tr\bigg\{\bigg( \sum_{i=1}^{D} D^{\ad}_i F_{ij}(x) \bigg)^2 \bigg\} + \calO(a^7) .
    \label{eq:Wact2}
\end{align}
Collecting Eqs.~\eqref{eq:Pact2} and \eqref{eq:Wact2},
\begin{equation}
\begin{split}
S_\lat 
& \to \frac{1}{2} \int \dd x_0 \dd^Dx \bigg[ \frac{1}{e_\lat^2}\frac{b}{a^{D-2}} \sum_{i=1}^{D}\Tr\ckakko{F_{0i}(x)^2} \\
&\quad + \frac{1}{g_\lat^2}\frac{a^{6-D}}{b} \sum_{j=1}^{D} \Tr\bigg\{ \bigg( \sum_{i=1}^{D} D^{\ad}_i F_{ij}(x) \bigg)^2 \bigg\} \bigg]
\end{split}
\end{equation}
as $a,b\to 0$.
This reproduces the continuum action \eqref{eq:action}. 

For completeness we outline 
the lattice discretization of other possible terms in the action in 
appendix \ref{sc:app2}.

\begin{figure}[t]
  \centering
  \includegraphics[width=.33\textwidth]{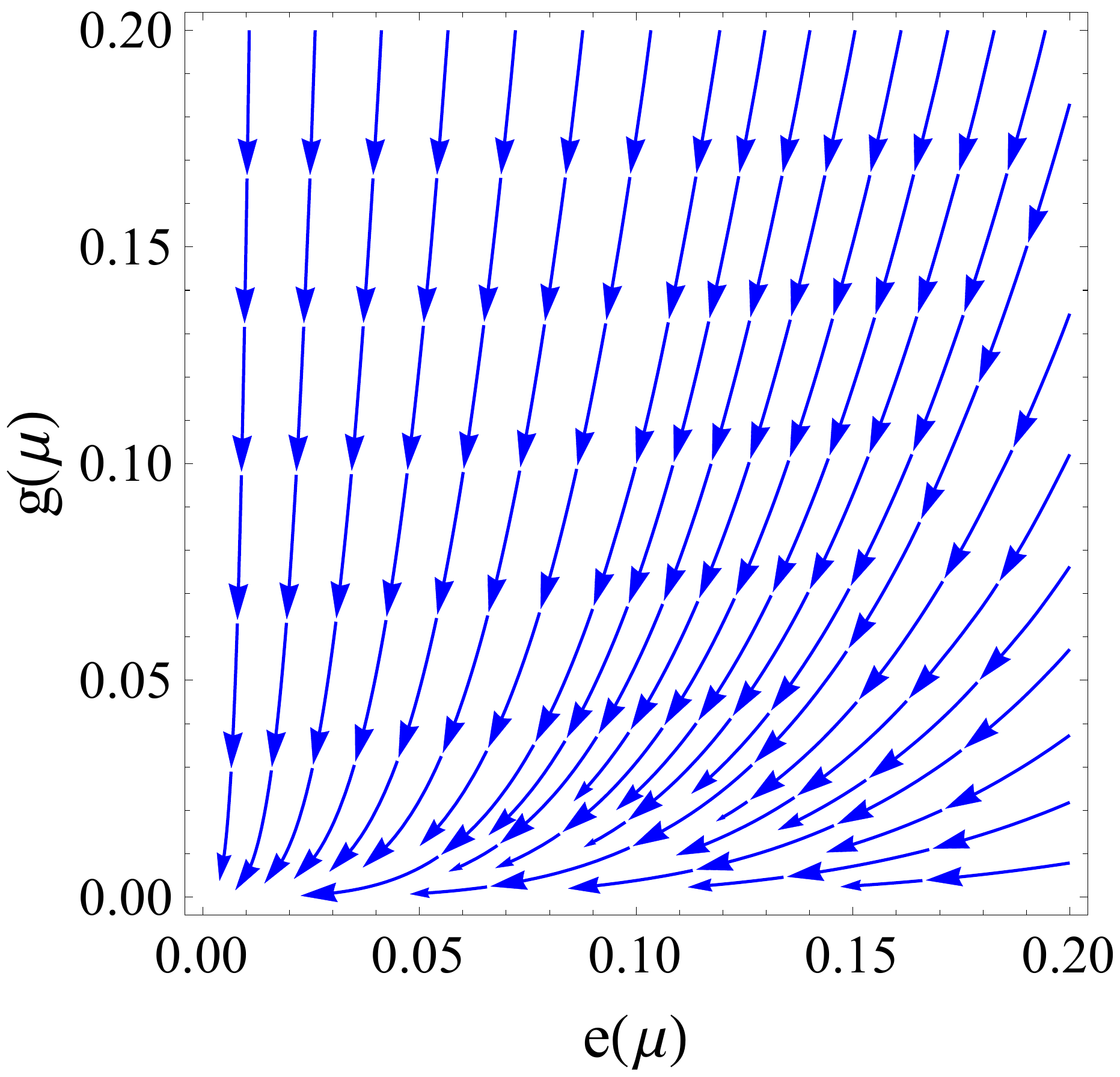}
  \vspace{-.5\baselineskip}
  \caption{\label{fg:egflow}
  The flow diagram of $e$ and $g$ for the $\SU(N)$ gauge group. 
  The origin is a UV fixed point. }
\end{figure}

Matching with the continuum action \eqref{eq:action} yields 
\begin{align}
  & \frac{1}{e^2} = \frac{1}{e_\lat^2}\frac{b}{a^{D-2}}
  \quad \text{and} \quad 
  \frac{1}{g^2} = \frac{1}{g_\lat^2}\frac{a^{6-D}}{b} \,. 
  \label{eq:u1scaling}
\end{align}
The two terms in Eq.~\eqref{eq:Slat} are of the same order \emph{only if} we take the limit $a,b\to 0$ with $b/a^2\sim \calO(1)$. 
Plugging this scaling into Eq.~\eqref{eq:u1scaling}, we find $e_\lat^2\sim e^2 a^{4-D}$ and $g_\lat^2\sim g^2 a^{4-D}$.   
Now, let us consider the continuum limit in each dimension:  
\begin{itemize}
  \item $D=2$ ($d=2+1$):~$e^2_\lat$, $g^2_\lat \propto a^2~\Rightarrow~$ $e_\lat$, $g_\lat \to 0$ ~
  with $e_\lat/g_\lat\sim \calO(1)$. 
  \item $D=3$ ($d=3+1$):~$e^2_\lat$, $g^2_\lat \propto a^1~\Rightarrow~$ $e_\lat$, $g_\lat \to 0$ ~
  with $e_\lat/g_\lat\sim \calO(1)$.  
  \item $D=4$ ($d=4+1$): $e^2_\lat$, $g^2_\lat \propto a^0~\Rightarrow~$ It is unclear how to take the continuum limit at tree level. 
\end{itemize}
This means that the continuum limit for $D=2$ and $3$ ($d=3$ and $4$) is reached trivially by sending $e_\lat$ and $g_\lat$ to $0$. 
However, $D=4$ ($d=5$) is the critical dimension where there is no scaling of the couplings 
at tree level.
In $D=4$, the one-loop $\beta$ functions \cite{Horava:2008jf} are given by
\begin{subequations}
\begin{align}
  \frac{\dd}{\dd\log \mu} e(\mu) & = - \frac{3}{2}C_2 e^2g + \cdots
  \\
  \frac{\dd}{\dd\log\mu} g(\mu) & = - \frac{35}{6}C_2 eg^2 + \cdots ,
\end{align}
\label{eq:egflow}
\end{subequations}
or, with $g_{\YM}\equiv \sqrt{eg}$ and $\lambda\equiv g/e$,
\begin{subequations}
\begin{align}
  \frac{\dd}{\dd\log \mu}g_\YM(\mu) & = - \frac{11}{3}C_2 g^3_\YM + \calO(g^5_\YM)
  \label{eq:betag}
  \\
  \frac{\dd}{\dd\log\mu}\lambda(\mu) & = - \frac{13}{3}C_2 g^2_{\YM}\lambda + \calO(g^4_{\YM}\lambda)\,,
  \label{eq:betalam}
\end{align}
\end{subequations}
where $C_2 \equiv N/(4\pi^2)$ for the gauge group $\SU(N)$. 
The theory is asymptotically free and therefore the continuum limit is achieved 
by sending both $g_\YM$ and $\lambda$ to $0$. 
Solving Eqs.~\eqref{eq:betag} and \eqref{eq:betalam} simultaneously, we find 
\begin{align}
  \lambda(\mu) \propto \big(g_{\YM}(\mu)\big)^{13/11}\,, ~~~~\text{i.e.,}~~~ g \propto e^{35/9}\,.
\end{align}
This scaling defines lines of constant physics in the weak-coupling region 
on the $(e,g)$ plane. 
The renormalization group flow of $e$ and $g$ is displayed in Fig.~\ref{fg:egflow}. 
(Since $C_2$ only enters the $\beta$ functions \eqref{eq:egflow} as a multiplicative 
factor, the flow pattern is the same for all $N\geq 2$.) 
Integrating Eq.~\eqref{eq:betag}, we encounter an infrared energy scale which survives the continuum limit: 
\begin{equation}
  \Lambda = \frac{1}{a}\exp\bigg(-\frac{24\pi^2}{11} \frac{1}{N g^2_\YM(\frac{1}{a})} \bigg)\,. 
\end{equation}
This is the phenomenon called dimensional transmutation.

The above formulation is straightforwardly applicable to Abelian gauge theories as well. 
The geometrical structure of the lattice action is the same, with 
$\SU(N)$ link variables replaced with $\U(1)$ link variables.  
However, the resultant compact $\U(1)$ gauge theory is not asymptotically free in $D=4$ $(d=5)$.

\section{Numerical simulation}
\label{sc:numerics}

We apply the above formulation to the lattice Monte Carlo simulation. 
The simulation can be done with standard algorithms in the lattice Yang-Mills theory.
In this work, we performed a simulation of the lattice Ho\v{r}ava-Lifshitz theory for the case of 
$\SU(N=3)$ gauge group.

First we examine the bulk phase structure on the $(e,g)$ plane.
We calculated the action density $s \equiv \langle S_\lat \rangle /N_\lat$ 
for various values of the lattice couplings defined as 
\begin{align}
  \beta_e \equiv \frac{2N}{e_{\lat}^2} ~~~\text{and}~~~ 
  \beta_g \equiv \frac{2N}{g_{\lat}^2}\,.
\end{align}
The lattice size is $N_{\lat} = 6^5$.
(We partially checked the volume independence of the action density on a $10^5$ lattice.) 
For isotropic couplings ($\beta_e=\beta_g\equiv\beta$), we find 
using standard analytical methods \cite{Creutz:1984mg} that 
the action density behaves as
\begin{subequations}
\begin{align}
  s &= D \beta + \calO(\beta^2) && (\beta \to 0), 
  \label{eq:sstrong}
  \\
  s &= (N^2-1)D/2 + \calO(1/\beta) && (\beta \to \infty),
  \label{eq:sweak}
\end{align}
\end{subequations}
respectively. This is useful in checking numerical data. 

\begin{figure}[t]
\includegraphics[scale=1.4]{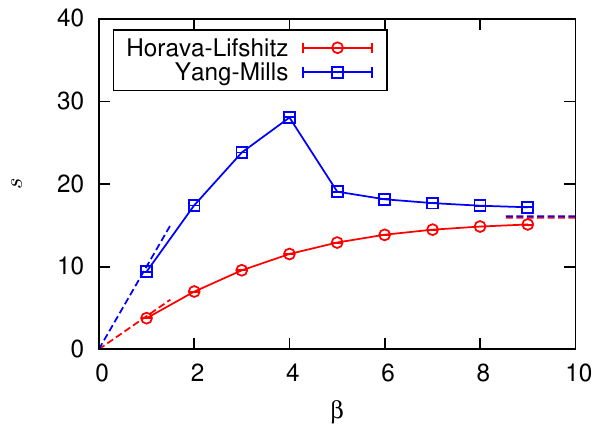}
\vspace{-2\baselineskip}
\caption{\label{fig1}
Action density with isotropic coupling constants 
$\beta\,(\,\equiv \beta_e = \beta_g)$.
The data of the Ho\v{r}ava-Lifshitz theory and the isotropic 
Yang-Mills theory on a $6^5$ lattice are plotted. 
The dashed lines are analytic results in the strong coupling limit 
($\beta \to 0$) and the weak coupling limit ($\beta \to \infty$). 
}
\end{figure}

\begin{figure}[t]
\includegraphics[scale=1.2]{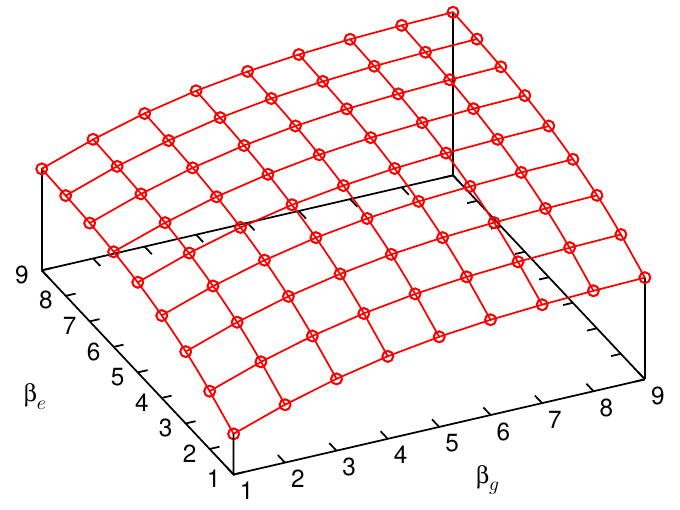}
\caption{\label{fig2}
Action density of the Ho\v{r}ava-Lifshitz theory as a function of $\beta_e$ and $\beta_g$.
The data on a $6^5$ lattice are plotted.
Statistical error bars are omitted.
}
\end{figure}

In Fig.~\ref{fig1}, we show the simulation results for isotropic couplings $\beta \equiv \beta_e = \beta_g$.
For comparison, we also show simulation results of the isotropic Yang-Mills theory in five dimensions.
As already known, there is a jump at $\beta = 4$--5 in the five-dimensional lattice Yang-Mills theory \cite{Itou:2014iya}. 
This jump indicates a bulk first-order phase transition from a confining phase to a deconfined phase. 
This bulk phase transition is a lattice artifact.
Its existence reflects the non-renormalizable nature of the lattice Yang-Mills theory in five dimensions.
On the other hand, there seems to be no phase transition in the Ho\v{r}ava-Lifshitz theory. 
In Fig.~\ref{fig1}, the dashed lines are asymptotics in the strong coupling limit (\ref{eq:sstrong}) and 
in the weak coupling limit (\ref{eq:sweak}) 
\footnote{In the strong coupling limit of the isotropic Yang-Mills theory  
in $d$ dimensions, \mbox{$s=d(d-1)\beta/2+\calO(\beta^2)$}. 
In the present case ($d=5$), \mbox{$s=10\beta+\calO(\beta^2)$}. 
In the weak coupling limit, the behavior of $s$ at leading order 
is the same as in the Ho\v{r}ava-Lifshitz theory and is given by \eqref{eq:sweak}.}.
The action density varies smoothly from the strong coupling limit to the weak coupling limit. 
As shown in Fig.~\ref{fig2}, there is no discontinuity in the region $1 \le \beta_e \le 9$ and $1 \le \beta_g \le 9$.
Thus, we can smoothly take the continuum limit of the lattice Ho\v{r}ava-Lifshitz theory.

\begin{figure}[t]
\includegraphics[scale=1.4]{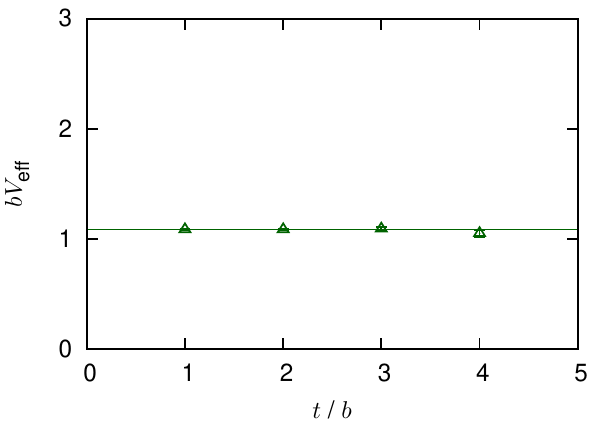}
\vspace{-\baselineskip}
\caption{\label{fig3}
Effective mass $V_{\rm eff}(t,x)$ at $x/a = 2$.
The data from simulations on a $10^5$ lattice 
with $\beta\,(\,\equiv \beta_e = \beta_g) = 9$ are plotted.
}
\end{figure}

Next we study a rectangular Wilson loop $W_{0i}$ lying in the $(x_0,x_i)$ plane.
The lattice size is $N_{\lat} = 10^5$.
The temporal Wilson loop may be interpreted as the infinite mass limit of a quark-antiquark system. 
(Although a Lifshitz-type fermion action admits various kinds of terms \cite{Anselmi:2008bq,Anselmi:2008bs,Dhar:2009dx,Bakas:2011uf}, this interpretation for the temporal Wilson loop should be correct provided that fermions couple to the temporal gauge field in a minimal way, 
as $\bar\psi\gamma_0 D_0\psi$.)
It gives the color singlet potential
\begin{equation}
 V(x) = - \lim_{t\to \infty} \frac{1}{t} \ln \langle W_{0i} (t,x) \rangle .
\end{equation}
In numerical simulations, the extrapolation to the limit $t \to \infty$ is done through a numerical fitting in a large but finite range of $t$.
To check the fit-range independence, we plot the effective mass $bV_{\rm eff}(t,x) = - \langle \ln \{ W_{0i} (t+b,x)/W_{0i} (t,x) \} \rangle$ in Fig.~\ref{fig3}.
The fit-range independence is clearly seen.

\begin{figure}[t]
\includegraphics[scale=1.4]{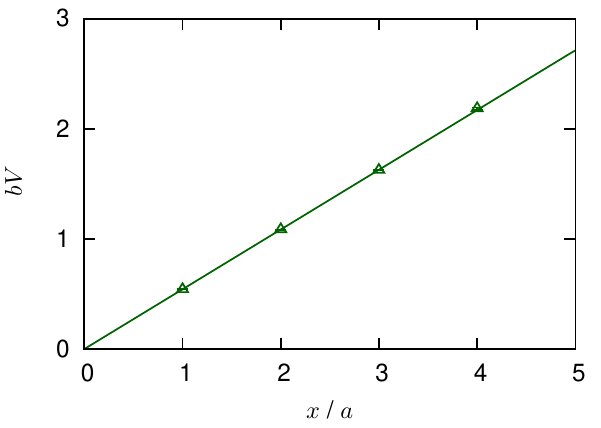}
\vspace{-\baselineskip}
\caption{\label{fig4}
Color singlet potential $V(x)$.
The data from simulations on a $10^5$ lattice 
with $\beta\,(\,\equiv \beta_e = \beta_g) = 9$ are plotted.
}
\end{figure}

In Fig.~\ref{fig4} we show numerical results of the color singlet potential.
The potential is linear.
Therefore the Ho\v{r}ava-Lifshitz theory is a confining theory.
We can analytically calculate the color singlet potential in two different limits:
(i) In the strong coupling limit, the strong coupling expansion is justified. At leading order, 
we can prove that the Wilson loop obeys an area law and thus the potential is linear. 
The proof is exactly the same as the famous proof in the Yang-Mills theory \cite{Wilson:1974sk} 
because the temporal component of the lattice action is given by the plaquettes $P_{0i}$ 
both in the Ho\v{r}ava-Lifshitz theory and in the Yang-Mills theory. 
(ii) In the short distance limit, the perturbative loop expansion is justified because the theory is asymptotically free.
Since the gluon propagator of $A_0$ is $\sim 1/p^2$, the perturbative 
one-gluon-exchange potential is $V(x) \sim \int \dd^D p \exp(-ipx)/p^2 \sim 1/x^2$.
However, this correction cannot be seen in Fig.~\ref{fig4}.
Its coefficient must be very small or zero.

We also measured the expectation values of spatial plaquettes $P_{ij}$ and 
spatial Wilson loops $W_{ij}$ and found them to be zero within errors.
This means in particular that the field strength $\Tr(F_{ij}^2)$ is not induced in the action, 
which is consistent with the renormalizability of the theory due to the detailed balance condition \cite{Horava:2008jf}.
They can be nonzero if spatial plaquettes or other deformation terms are added to the action.

\section{Summary}
\label{sc:summary}

We proposed a lattice formulation of the Ho\v{r}ava-Lifshitz-type gauge theory. 
For a non-Abelian gauge group they are asymptotically free even in five dimensions. 
We performed the first Monte Carlo simulation of this theory on a lattice for the $\SU(3)$ gauge group. Numerical results suggest that the continuum limit can be taken smoothly, in contrast to the ordinary Yang-Mills theory in 
five dimensions which is beset with a bulk phase transition. 
Using the present framework one can study various nonperturbative aspects of 
the Ho\v{r}ava-Lifshitz-type gauge theories by means of numerical lattice simulations. 
For example, it is straightforward to compactify a temporal or spatial direction and study possible center symmetry breaking.  
Of course one can perform simulations for other gauge groups and in other spacetime dimensions. 
Lattice simulations may also be performed with additional terms in the action, such as $\Tr(F_{ij}^2)$, $\Tr(F_{ij}F_{jk}F_{ki})$, and 
$\Tr\big\{ (D_i^{\ad}F_{jk})(D_i^{\ad}F_{jk}) \big\}$, as discussed 
in appendix \ref{sc:app2}. The interplay of these terms is an interesting subject. 
A more ambitious generalization is to include fermions coupled to the gauge field 
and study spontaneous chiral symmetry breaking.
These issues are left for future works.

\acknowledgments
TK was supported by the RIKEN iTHES Project and JSPS KAKENHI Grants Number 25887014.
The numerical simulations were performed by using the RIKEN Integrated Cluster of Clusters (RICC) facility.

\appendix 
\section{Classical continuum limit}
\label{sc:continuum}

It has been known from \cite[Eq.\,(16)]{Luscher:1984xn} that 
a spatial plaquette in the naive continuum limit $a\to 0$ becomes
\begin{align}
  & P_{ij}(x) \equiv  
  U_i(x)U_{j}(x+a\hat{i})U_i(x+a\hat{j})^\dagger U_j(x)^\dagger
  \notag
  \\
  & = \exp \Big( ia^2F_{ij}(x) + 
  \frac{i}{2}a^3 (D^{\ad}_{i}+D^{\ad}_{j})F_{ij}(x) + \calO(a^4) \Big)\,. 
  \label{eq:Pspace}
\end{align}
Because a twisted $2\times 1$ Wilson loop is a product of 
two neighboring spatial plaquettes, we get
\begin{align}
  & \quad T_{ij}(x) 
  \notag
  \\
  & = 
  \exp \Big( ia^2F_{ij}(x) + \frac{i}{2}a^3 (D^{\ad}_{i}+D^{\ad}_{j})F_{ij}(x) + \calO(a^4) \Big)
  \notag
  \\
  & ~\times \exp \Big(\! - ia^2F_{ij}(x) 
  - \frac{i}{2}a^3 ( - D^{\ad}_{i} + D^{\ad}_{j} )F_{ij}(x) + \calO(a^4) \Big)
  \notag
  \\
  & = \exp \big( ia^3 D^{\ad}_{i} F_{ij}(x) + \calO(a^4) \big)\,, 
\end{align}
which proves \eqref{eq:Tij}.

\section{More general lattice action}
\label{sc:app2}

Besides $\Tr\{ (D_i^{\ad}F_{ik})(D_j^{\ad}F_{jk}) \}$, there are many other terms 
that could have been added to the action \eqref{eq:action}.  
In this appendix we discuss how to discretize them on a lattice.  

Firstly, the term $\Tr\mkakko{ F_{ij}F_{jk}F_{ki} }$ can be realized on a lattice 
as follows.  Let us consider 
\begin{align}
  \Tr\ckakko{ \mkakko{\1-P_{ij}(x)}\mkakko{\1-P_{jk}(x)}\mkakko{\1-P_{ki}(x)} } \,. 
  \label{eq:FFF}
\end{align}
This expression is manifestly gauge invariant. By plugging in 
\eqref{eq:Pspace} for each $P$ and expanding in powers of $a$ we get
\begin{align}
  \eqref{eq:FFF} & = i a^6 \Tr\ckakko{ F_{ij}(x) F_{jk}(x) F_{ki}(x) } + \calO(a^7)\,, 
\end{align}
which is the desired term.

The second term of our interest is $\Tr\ckakko{ D_k^{\ad}F_{ij}(x) D_k^{\ad}F_{ij}(x) }$. 
The case with $k=i$ or $k=j$ follows from $T_{ij}(x)$ as given in \eqref{eq:Tij}, 
so it is enough to assume here that $i,j$ and $k$ are distinct from each other, 
which requires $D\geq 3$.  

\begin{figure}[t]
  \centering \includegraphics[width=.7\columnwidth]{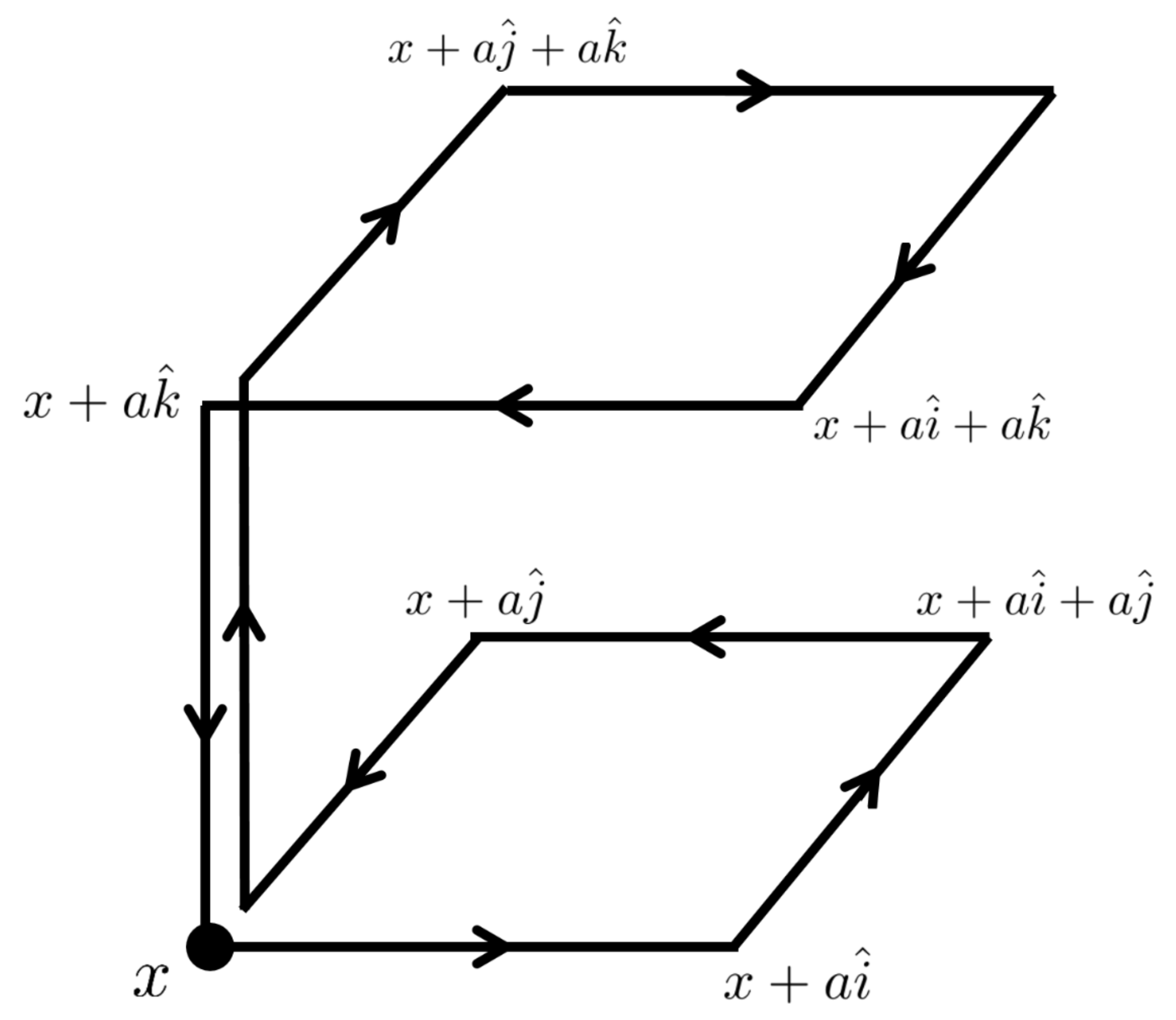}
  \caption{\label{fg:Wloop2}  A Wilson loop on a lattice which reproduces 
  $\Tr\{ (D_k^{\ad}F_{ij})(D_k^{\ad}F_{ij}) \}$ in the continuum limit. 
  }
\end{figure}

Let us start from a Wilson loop $W_{ijk}(x)$ 
shown in Fig.~\ref{fg:Wloop2}: 
\begin{align}
  W_{ijk}(x) & 
  \equiv P_{ij}(x)U_k(x)P_{ij}(x+a\hat{k})^\dagger U_k(x)^\dagger 
  \notag
  \\
  & = \ee^{a^2\PP_1}\ee^{iaA_k(x)+\calO(a^2)}
  \ee^{-a^2\PP_2}\ee^{-iaA_k(x)+\calO(a^2)}\,,
  \notag
\end{align}
where from \eqref{eq:Pspace} 
\begin{align*}
  \PP_1 & \equiv 
  iF_{ij}(x) + \frac{i}{2}a (D^{\ad}_{i}+D^{\ad}_{j})F_{ij}(x) + \calO(a^2)\,,
  \\
  \PP_2 & \equiv 
  iF_{ij}(x+a\hat{k}) + 
  \frac{i}{2}a (D^{\ad}_{i}+D^{\ad}_{j})F_{ij}(x+a\hat{k}) + \calO(a^2)
  \\
  & = \PP_1 + ia \der_k F_{ij}(x) + \calO(a^2)\,. 
\end{align*}
Using the BCH formula,  
\begin{align}
  W_{ijk}(x) & = \exp\Big( 
    a^2(\PP_1-\PP_2) - ia^3 [A_k(x),\PP_2] + \calO(a^4)
  \Big)
  \notag
  \\
  & = \exp\big(  \!-ia^3 D_k^\ad F_{ij}(x)+\calO(a^4)  \big)\,,
\end{align}
so that
\begin{align}
  \mathrm{Re}\Tr\{ \1 - W_{ijk}(x) \} & = 
  \frac{1}{2}a^6 \Tr\ckakko{ D_k^{\ad}F_{ij}(x) D_k^{\ad}F_{ij}(x) }
  \notag
  \\
  & \quad + \calO(a^7)\,.
\end{align}
However, it has been known from \cite[Eq.\,(2.10)]{Weisz:1982zw} that 
$\Tr\ckakko{ (D_i^{\ad}F_{ik})(D_j^{\ad}F_{jk}) }$, 
$\Tr\ckakko{ F_{ij}(x) F_{jk}(x) F_{ki}(x) }$ and $\Tr\ckakko{ D_k^{\ad}F_{ij}(x) D_k^{\ad}F_{ij}(x) }$ 
are linearly dependent, up to a total derivative. Thus it is sufficient to keep only 
two of them in the action. 

The lattice actions for other possible terms like 
$\epsilon_{jklm}\Tr\ckakko{ D_i^{\ad}F_{jk}(x) D_i^{\ad}F_{lm}(x) }$ 
(for $D=4$) can be worked out along similar lines.

\bibliography{HoravaYM}
\end{document}